\documentclass[11pt]{article}
\usepackage[margin=1in]{geometry}
\usepackage{graphicx}
\usepackage{authblk}
\usepackage[sc]{mathpazo}
\usepackage[cm]{fullpage}
\usepackage{hyperref}

\def\beq{\begin{equation}}
\def\eeq#1{\label{#1}\end{equation}}
\def\eeqn{\end{equation}}

\def\beqa{\begin{eqnarray}}
\def\eeqa#1{\label{#1}\end{eqnarray}}
\def\eeqan{\end{eqnarray}}

\let\bar=\overbar

\def\Dslash{\not{\hbox{\kern-4pt $D$}}}
\def\dslash{\not{\hbox{\kern-2pt $\del$}}}

\def\msb{{\bar{\ssstyle M \kern -1pt S}}}

\def\Conference{\vspace{4mm}\begin{raggedright} {\normalsize {\it Talk presented at the 2019 Meeting of the Division of Particles and Fields of the American Physical Society (DPF2019), July 29--August 2, 2019, Northeastern University, Boston, C1907293.} } \end{raggedright}\vspace{4mm}}

\newcommand{\PTm}{\ensuremath{{p}_\mathrm{T}\hspace{-1.02em}/\kern 0.5em}\xspace}

\newcommand{\ETslash}{\ensuremath{E_{\mathrm{T}}\hspace{-1.1em}/\kern0.45em}\xspace}

\newcommand{\ptvecmiss}{\ensuremath{{\vec p}_{\mathrm{T}}^{\kern1pt\text{miss}}}\xspace}

\newcommand{\rpv}{\ensuremath{\rlap{\kern.2em/}R}\xspace}

\begin{document}

%
%

\title{\textbf{End-to-end particle and event identification at the Large Hadron Collider with CMS Open Data}}

\author[1]{M. Andrews}
\author[1]{J. Alison}
\author[1,2]{S. An}
\author[1]{P. Bryant}
\author[3]{B. Burkle}
\author[4]{S. Gleyzer}
\author[3]{M. Narain}
\author[1]{M. Paulini}
\author[5]{B. Poczos}
\author[3]{E. Usai}

\affil[1]{Department of Physics, Carnegie Mellon University, Pittsburgh, USA}
\affil[2]{CERN, Geneva, Switzerland}
\affil[3]{Department of Physics, Brown University, Providence, USA}
\affil[4]{Department of Physics, University of Alabama, Tuscaloosa, USA}
\affil[5]{Machine Learning Department, Carnegie Mellon University, Pittsburgh, USA}

\maketitle
\abstract{From particle identification to the discovery of the Higgs boson, deep learning algorithms have become an increasingly important tool for data analysis at the Large Hadron Collider (LHC).
We present an innovative end-to-end deep learning approach for jet identification at the Compact Muon Solenoid (CMS) experiment at the LHC. The method combines deep neural networks with low-level detector information, such as calorimeter energy deposits and tracking information, to build a discriminator to identify different particle species. Using two physics examples as references: electron vs. photon discrimination and quark vs. gluon discrimination, we demonstrate the performance of the end-to-end approach on simulated events with full detector geometry as available in the CMS Open Data. We also offer insights into the importance of the information extracted from various sub-detectors and describe how end-to-end techniques can be extended to event-level classification using information from the whole CMS detector.}

\Conference
%
%

\section{Introduction}

The analysis of data collected at the LHC often involves the classification of hadronic jets and particles in collision events. The CMS experiment \cite{cms} uses the particle flow algorithm \cite{pf} that combines raw data from different subdetectors to identify different classes of particles.   
The particle-flow objects are then clustered into jets and used as an input to algorithms aimed at distinguishing different flavours of the jet (light quark, gluon, c quark, b quark, and top quark).
Recently, jet classification algorithms based on deep neural networks have been introduced \cite{deepjet, deepflavour}, achieving the current state-of-the-art performance in jet classification tasks, often using simplified detector models.

\section{The End-to-end approach}
\label{approach}

This work presents an innovative approach to the classification of jets and particles that instead uses low-level detector information. This approach is called end-to-end (E2E) classification and in this work is applied to develop a discriminator between light quark and gluon-initiated jets \cite{e2e2}. 

The E2E approach consists of building a convolutional neural network (CNN)-based jet or event discriminator by using the low-level detector inputs directly. We use an image-based approach where the information from each subdetector is represented by one or more image maps and combined with those from other subdetectors to form a single multi-layer image. The pixel coordinates of the image are then mapped to the discretized detector coordinates expressed in azimuthal angle $\phi$ the pseudorapidity $\eta$.

The work presented here uses simulated datasets from the CMS Open Data repository \cite{opendata}, which are essential to ensuring a high-fidelity representation of the detector. New releases of the CMS Open Data portal continue to be published that specifically target machine learning (ML) studies \cite{od1,od2}, including end-to-end-oriented workflows.
For instance, tools and instructions for reproducing the full simulation chain of CMS are available allowing access to any kind of low-level detector information one may need.

The first step to the construction of the detector images is to generate an image for the electromagnetic calorimeter (ECAL). For this subdetector, each pixel in the image corresponds to a crystal of ECAL. The intensity of each pixel is proportional to the energy measured in the corresponding crystal. Other data attributes like timing of the energy deposit are also available, though this did not yield any further improvement for the classification tasks considered in this work. This may potentially change, however, with the improved timing capabilities of the Phase 2 upgrades of the CMS detector.

The second image considered contains information from the hadronic calorimeter (HCAL). In this case, each pixel of the image corresponds to one of the calorimeter towers. Similarly, the intensity of each pixel contains the energy collected from each tower. By construction, a 5x5 window of ECAL pixels corresponds to a single HCAL pixel.

\begin{figure}[htb]
\centering
\includegraphics[height=3in]{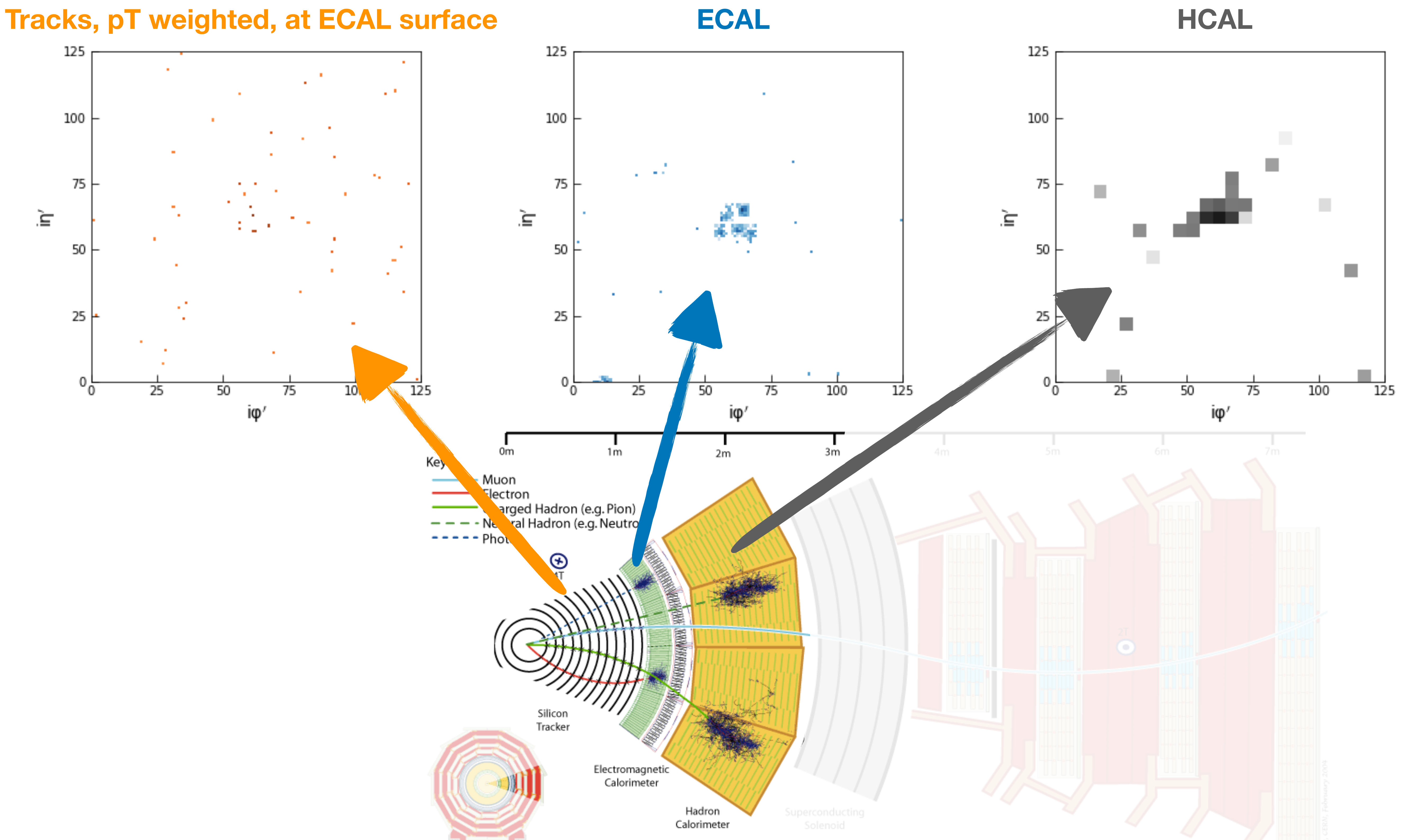}
\caption{E2E images for the electromagnetic calorimeter, the hadronic calorimeter and the tracker of the CMS detector.}
\label{fig:e2e4}
\end{figure}

Finally, for the tracking system, reconstructed tracks are used. The position of the track is propagated to the ECAL surface. The $(\eta,\phi)$ coordinates of the track position at the ECAL surface are discretized on a grid with the same granularity as the ECAL's. The intensity of each pixel corresponds to the transverse momentum of the track. Figure \ref{fig:e2e4} shows the E2E images for the three subdetectors detectors described above.

One of the problems of generating detector images is that the some endcap detectors use a different coordinate system. For example, while the HCAL endcap sensor elements are discretized in ($\eta$, $\phi$) coordinates, the ECAL endcaps are discretized in Cartesian coordinates ($x,y$).
In order to obtain a single image from the barrel and the endcap regions of a subdetector, the images need to be stitched together in the same coordinate system. This is done by morphing the endcap images, making a coordinate transform from ($x,y$) to ($\eta$, $\phi$) then stitching them to the edges of the barrel image. 

The full-detector E2E images can be directly fed to a CNN for training an event-wide discriminator. On the other hand, smaller images can be cropped from the full-detector around region of interests, such as jets or photons.
In the case of jet discriminators we crop an 125x125 pixel image around the jet.


\section{Previous work}

This section briefly summarizes the previous applications of the E2E framework \cite{e2e1}. 
The first application of E2E was the construction of a discriminator between photons and electrons. This application used only ECAL energy images. The E2E images were fed to a ResNet-15 \cite{DBLP:conf/cvpr/2016} network obtaining an area under the curve of the Receiver Operating Charactersitic (ROC AUC) score of 0.788. Additionally, a di-photon vs. di-electron discriminator was built using similarly constructed detector images and a ResNet-23 network, obtaining a ROC AUC of 0.997.

The E2E framework was also used to distinguish the Higgs boson to di-photon decay from non-resonant di-photon production and misidentified photons. This application used ECAL, HCAL and track images. One of the challenges in this particular application was the correlation of the discriminator output with the mass of the Higgs boson, which is an undesirable feature in physics analyses. The discriminator was de-correlated from the diphoton mass through the use of an additional loss penalty proportional to the learned mass correlation as measured by the Cramer-Von Mises metric \cite{CVM}. 
The de-correlated discriminator yielded a ROC AUC score of 0.77 in distinguishing $H\rightarrow \gamma\gamma$ from non-resonant $\gamma\gamma$ production, 0.95 for $H\rightarrow \gamma\gamma$ vs. $\gamma+$jet (misidentified photon), and 0.81 for $H$ vs. non-resonant $\gamma\gamma$ and $\gamma+$jet together. 

\section{Quark- vs. gluon-jet discriminator}

\begin{figure}[htb]
\centering
\includegraphics[height=3in]{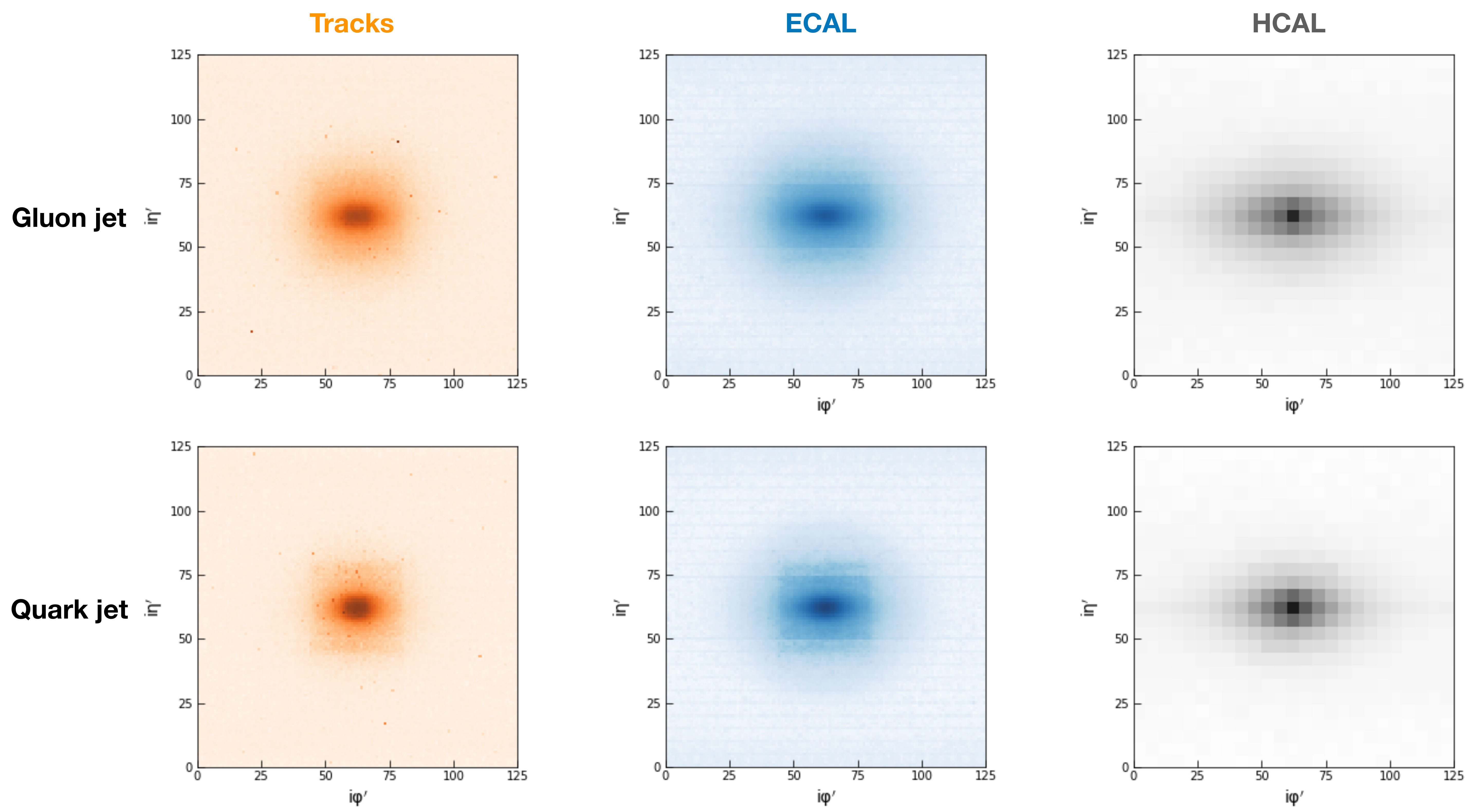}
\caption{E2E detector images for quark and gluon jets, averaged over many events}
\label{fig:e2e1}
\end{figure}

Similarly to the $H\rightarrow \gamma\gamma$ case, the E2E quark vs. gluon jet classifier uses three image channels from the ECAL, HCAL, and track momentum. Figure \ref{fig:e2e1} shows the three E2E detector images for quark and gluon jets, averaged over 70k jets. 

To understand the relevance of the information contained in the different image layers, the ResNet-15 network is first trained using images from a single subdetector alone. The tracker layer shows the highest ROC AUC score at 0.782, followed by ECAL (0.760), and finally HCAL (0.682). Following this, we can assess how much of the information in the different layers is redundant by training on different combinations of the subdetector layers. The combination of ECAL and HCAL images has a ROC AUC of 0.781, slightly below using the track layer alone. On the other hand, adding ECAL to just the track layer brings a considerable improvement to the ROC AUC at 0.804. Finally, using all three subdetectors together brings only a modest improvement (0.808).

\begin{figure}[htb]
\centering
\includegraphics[height=3.5in,angle=90]{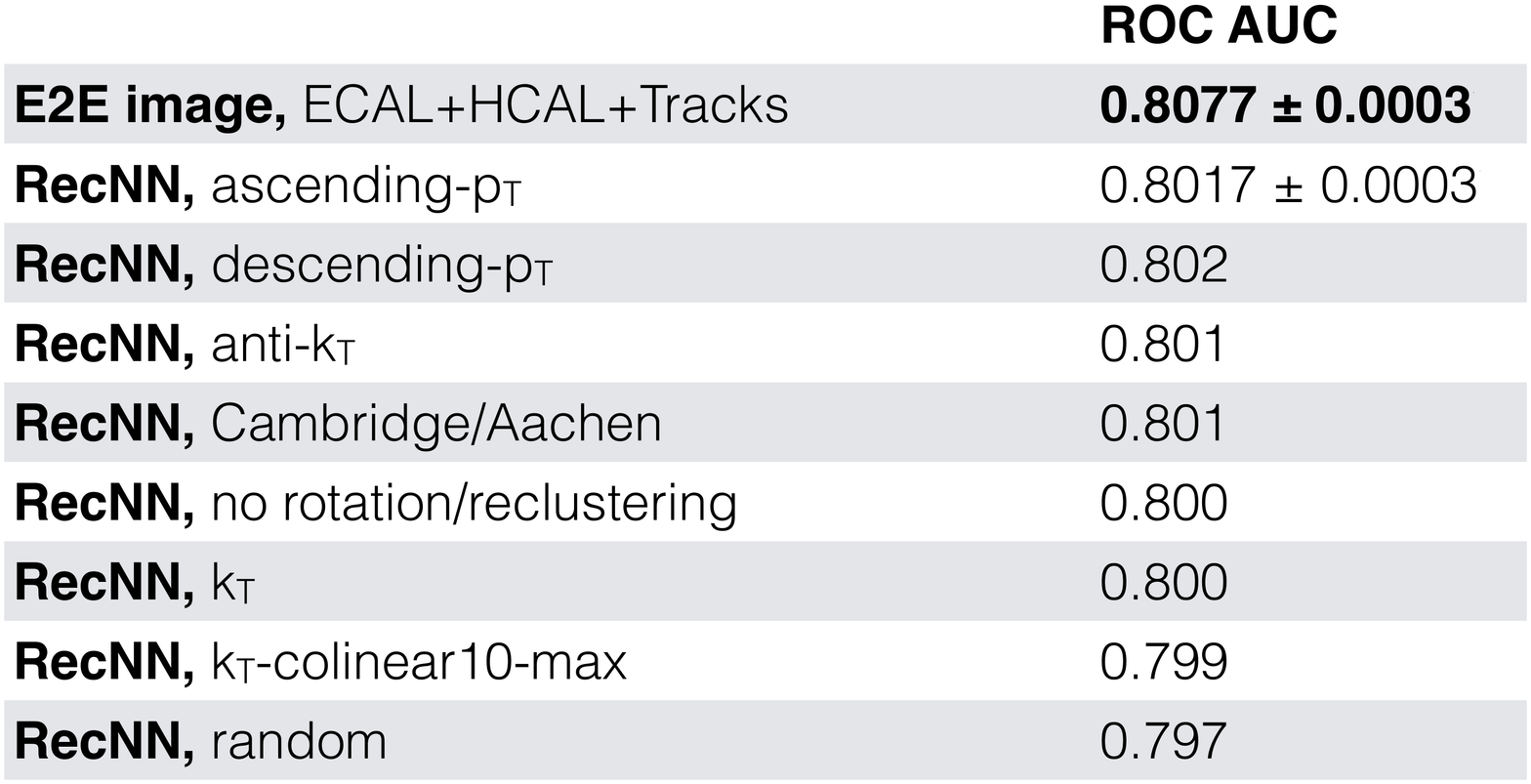}
\caption{Performance comparison of the E2E jet classifier vs. RecNN. *Represent ROC AUC mean $\pm$ standard deviation over five training runs.}
\label{fig:e2e2}
\end{figure}

The performance of the jet classifier is then benchmarked against the QCD-aware recursive neural network (RecNN) approach outlined in \cite{kyle} and adapted to the quark vs. gluon jet case in \cite{taoli}. Figure \ref{fig:e2e2} shows a comparison between the E2E and RecNN quark vs. gluon discriminators using the same jets selected from the CMS Open Data dataset. The E2E jet classifier thus has performance comparable to the existing state-of-the-art, and performs significantly better than traditional jet images.

The final step consists of building a di-jet qq vs gg discriminator.
Three different scenarios are considered. In scenario A, we crop images around the two jets. In scenario B, the four-momenta of the two jets are included with the images. Lastly, in scenario C, the full-detector E2E images are used. Scenario C has the best performance with a ROC AUC of 0.889, followed by scenario B (0.878) and a (0.876).  
Approaches A and B have a very similar performance, while using a full detector image gives the best result.

\section{Recent developments}


Recent developments in the E2E framework goes in the direction of using more and lower-level detector information, particularly from the tracking system. Different image channels are added for different parameters of the reconstructed tracks. In addition to the transverse momentum, new layers are added that encode the transverse and longitudinal projections of the impact parameter of the track, dz and d0, computed with respect to the primary vertex of the interaction. The images for these new channels have the same pixel distribution as the transverse momentum image, but each pixel is assigned a different intensity, depending on the track parameter represented. These additional parameters allow the encoding of information from pileup vertices and secondary vertices coming from the hadronization of b quarks. 

\begin{figure}[htb]
\centering
\includegraphics[height=3in]{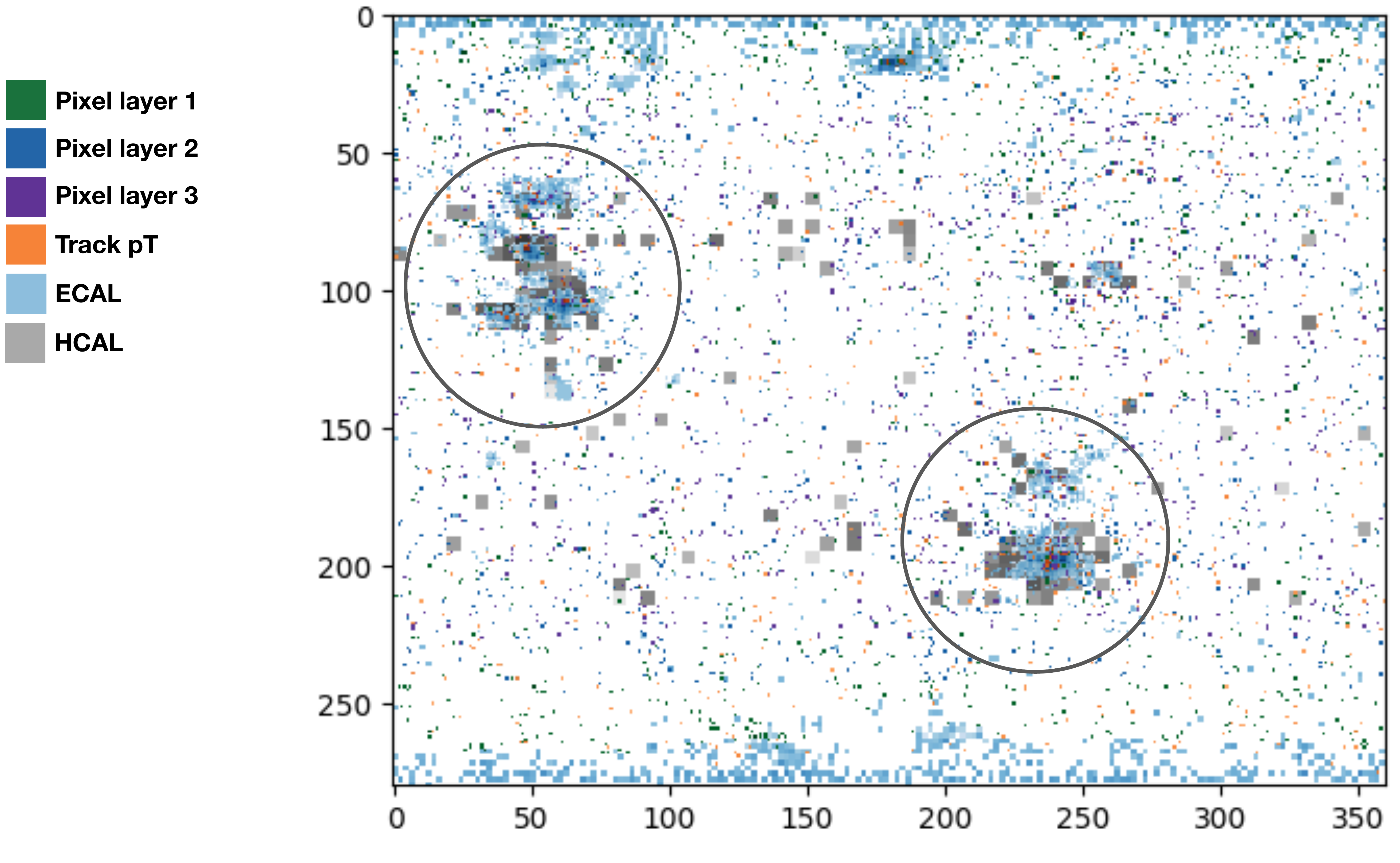}
\caption{E2E detector images including new tracker layers.}
\label{fig:e2e3}
\end{figure}

Another idea still involves using even lower-level tracker information. The CMS tracking system includes a three-layer pixel detector placed very close to the beampipe. In particular, each of the three pixel detector layers can each be transformed in an image layer. Each pixel can be either 0 or 1 depending on the presence of a hit in the corresponding position.	The resolution of the images could also be increased to fully take advantage of the higher resolution of the tracking system. Figure \ref{fig:e2e3} shows a full-detector E2E image with these three pixel layers in addition to the original ECAL, HCAL, and track momentum layers.

\section{Conclusion}

The E2E approach can be used for particle and jet identification. We show that the algorithm can learn particle kinematics and shower shapes.
We adapted the E2E framework to build a quark- vs. gluon-initiated jet discriminator. The performance of the algorithm is competitive to the current state-of-the-art for this task. 
Work is currently ongoing to extend this approach to include even lower-level tracker information and other classification tasks such as for top vs. light-flavour jet discrimination.

\bibliographystyle{lucas_unsrt}

\bibliography{bibliography}{}

\end{document}